%% file: main.tex
\pdfoutput=1

\documentclass[sigconf,nonacm]{acmart}

\usepackage{graphicx}

\usepackage{booktabs}

\usepackage{tabularx}
\usepackage{multirow}

\usepackage{subcaption}

\usepackage{seqsplit}

\usepackage{makecell}

\usepackage{verbatim}

\usepackage{amsmath}

\usepackage[nameinlink]{cleveref}

\usepackage{tabularray}

\usepackage{textcomp}

\usepackage{pifont}
\usepackage{xspace}
\usepackage{adjustbox}

\usepackage{xurl}

\usepackage[nolist]{acronym}

\usepackage{enumitem}

\usepackage{listings}
\definecolor{codegreen}{rgb}{0,0.6,0}
\definecolor{codegray}{rgb}{0.5,0.5,0.5}
\definecolor{codepurple}{rgb}{0.58,0,0.82}
\definecolor{backcolour}{rgb}{0.95,0.95,0.92}

\lstdefinestyle{mystyle}{
    backgroundcolor=\color{backcolour},   
    commentstyle=\color{codegreen},
    keywordstyle=\color{magenta},
    numberstyle=\tiny\color{codegray},
    stringstyle=\color{codepurple},
    basicstyle=\ttfamily\footnotesize,
    breakatwhitespace=false,         
    breaklines=true,                 
    captionpos=b,                    
    keepspaces=true,                 
    numbers=left,                
    numbersep=5pt,                  
    showspaces=false,                
    showstringspaces=false,
    showtabs=false,                  
    tabsize=2,
    frame=single,
    framesep=2pt,
    framexleftmargin=8pt,
    xleftmargin=0pt,
    xrightmargin=0pt
}

\crefname{lstlisting}{listing}{listings}
\Crefname{lstlisting}{Listing}{Listings}

\usepackage[framemethod=TikZ]{mdframed}
\global\mdfdefinestyle{insightstyle}{
	outerlinewidth=0.5pt,
	backgroundcolor=gray!10,
	linecolor=gray,
	innertopmargin=6pt,
	innerbottommargin=6pt,
	innerrightmargin=7pt,
	innerleftmargin=6pt
}
\global\mdfdefinestyle{recommendationstyle}{
	backgroundcolor=gray!10,
	linecolor=gray,
	innerrightmargin=7pt,
	innerleftmargin=6pt
}

\microtypecontext{spacing=nonfrench}
\begin{document}

\title{A Case Study of a Phishing Cluster Targeting German Universities}

\author{Raphael Springer}
\orcid{0009-0008-8298-7895}
\affiliation{
  \institution{Westphalian University of Applied Sciences}
  \department{Institute for Internet Security}
  \city{Gelsenkirchen}
  \state{}
  \country{Germany}
}
\email{springer@internet-sicherheit.de}

\author{Alexander Schmitz}
\orcid{0009-0006-0514-9535}
\affiliation{
  \institution{Westphalian University of Applied Sciences}
  \department{Institute for Internet Security}
  \city{Gelsenkirchen}
  \state{}
  \country{Germany}
}
\email{schmitz@internet-sicherheit.de}

\author{Joel Taddey}
% \orcid{}
\affiliation{
  \institution{aDvens GmbH}
  \city{Munich}
  \state{}
  \country{Germany}
}
\email{joel.taddey@advens.com}

\author{Gerhard Senkowski}
\orcid{0009-0009-9994-3925}
\affiliation{
  \institution{Westphalian University of Applied Sciences}
  \department{Institute for Internet Security}
  \city{Gelsenkirchen}
  \state{}
  \country{Germany}
}
\email{senkowski@internet-sicherheit.de}

\author{Artur Leinweber}
\orcid{0009-0001-7623-1038}
\affiliation{
  \institution{Westphalian University of Applied Sciences}
  \department{Institute for Internet Security}
  \city{Gelsenkirchen}
  \state{}
  \country{Germany}
}
\email{leinweber@internet-sicherheit.de}

\author{Christian Dietrich}
\orcid{0009-0001-5523-4467}
\affiliation{
  \institution{Westphalian University of Applied Sciences}
  \department{Institute for Internet Security}
  \city{Gelsenkirchen}
  \state{}
  \country{Germany}
}
\email{dietrich@internet-sicherheit.de}

\input{abstract}

\maketitle

\input{acronyms}
\input{introduction}
\input{related-work}
\input{cctgu}
\input{phishing-analysis}
\input{limitations}
\input{conclusion}

% \section*{Code Artifacts}

\bibliographystyle{plain}
\bibliography{bibliography}

\end{document}

%% file: abstract.tex
\begin{abstract}
In light of recent ransomware attacks, this study analyzes and measures the impact of cyberattacks against academic institutions in German-speaking areas.
First, we provide an overview of 25 confirmed cyberattacks against universities in German-speaking areas between February 2022 and February 2023.
In this time frame, we find that 5\% of all German universities have experienced a confirmed cyberattack with a measurable impact.
Second, we document victimology and \ac{TTP} of a phishing activity cluster spanning more than three years and showing a particular focus on universities in German-speaking areas.
Of the 93~targeted universities, 58~(62\%) are German, representing a significant fraction with approximately 14\%~of the total number of universities (423) in Germany.
A notable technique is that the phishing emails are sent via university infrastructure, using valid, compromised credentials of university-associated personnel and involving university \ac{VPN} tunnels.
\end{abstract}

%% file: acronyms.tex
\begin{acronym}
    \acro{TTP}{Tactics, Techniques, and Procedures}
    \acro{VPN}{Virtual Private Network}
    \acro{HAW}{Hochschule für Angewandte Wissenschaften}
    \acro{CISA}{Cybersecurity and Infrastructure Security Agency}
    \acro{FBI}{Federal Bureau of Investigation}
    \acro{U.S.}{United States}
    \acro{APT}{Advanced Persistent Threat}
    \acro{IRGC‑IO}{Iran’s Islamic Revolutionary Guard Corps–Intelligence Organization}
    \acro{NRW}{North Rhine-Westphalia}
    \acro{DDoS}{Distributed Denial of Service}
    \acro{FH}{Fachhoschule}
    \acro{HTW}{Hochschule für Technik und Wirtschaft}
    \acro{CSV}{Comma-Separated Values}
    \acro{URL}{Uniform Resource Locator}
    \acro{MTA}{Mail Transfer Agent}
    \acro{IP}{Internet Protocol}
    \acro{RWTH}{Rheinisch-Westfälische Technische Hochschule}
    \acro{UX}{User Experience}
    \acro{UI}{User Interface}
    \acro{2FA}{two-factor authentication}
\end{acronym}

%% file: introduction.tex
\section{Introduction}
In the last few years, several universities and academic institutions in Germany have been compromised through cyberattacks. 
For example, University Duisburg-Essen was compromised in November 2022 and has suffered a serious operational impact from a ransomware infection, claimed by the threat actor Vice Society~\cite{Bleeping_Computer_UDE_2023, WDR_DUE_3M_2023}.
The institution refused to pay.
Even three months later, several systems were still being rebuilt.
In addition, the victim likely suffered from a loss of reputation when exfiltrated information was published on a dedicated leak site in January 2023.
Similarly, \ac{HAW} Hamburg experienced a ransomware attack in December 2022, leading to data exfiltration and encryption of systems requiring a password reset and a laborious rebuild of systems~\cite{Tagesschau_HAW_Vice_2023}.
The Federal State Government of the State of North Rhine-Westphalia, Germany's most populated federal state, confirmed that all of its 77 universities have been targeted by some form of cyberattack attempt between 2018 and 2023~\cite{NRWAnfrage}.
While the financial damage is often difficult to assess, in a few cases, such figures highlight a significant financial impact. 
For example,
Hochschule Heilbronn was compromised in October 2022, leading to an estimated total of EUR~1~million in financial damage~\cite{Heilbronn_1_Mio_2023}.

In September 2022, U.S. law enforcement (i.e., the \ac{FBI}) and other security agencies (e.g., \ac{CISA}) published a joint advisory stating they recently observed the Vice Society actor disproportionately targeting the education sector with ransomware attacks~\cite{CISA_2022}.
This indicates a particular targeting of academic institutions, possibly with a \ac{U.S.} focus.
Studies find that some educational institutions paid the ransom~\cite{comparitech_ransomware_education_2023, sophos_ransomware_education_2023}, sometimes reporting that the educational sector exhibits one of the highest rates of ransom payment with more than half of the impacted higher educational organizations paying.

However, it remains unclear whether academia is particularly targeted in Germany.
Since most German universities are publicly funded, paying the ransom in case of a ransomware attack most likely imposes a significant legal obstacle.
For financially motivated threat actors, attacks against German universities are therefore unlikely to be considered promising and lucrative.
Occasionally, threat actors aim not merely for immediate financial gain but rather for long-term espionage, intellectual property theft, and influence over emerging technologies.
A recent prominent example emerged in mid-June 2025: Researchers uncovered a spear-phishing campaign run by Educated Manticore~\cite{educated_manticore_2025}, an Iranian-linked \ac{APT} group whose activities align with APT42 or Charming Kitten, believed to be connected to \ac{IRGC‑IO}. 
Targets included high-profile academics in technology and cybersecurity, as well as Israeli journalists, underscoring a strategic focus on information assets within academia.
In this study, we aim to find out more about cyberattacks and phishing activities targeting universities in particular by cyber threat actors with a focus on the German-speaking areas.

First, we provide a timeline of 25~confirmed cyberattacks against universities in German-speaking areas over the course of one year.
For each incident, we attempt to categorize the impact of the attack, distinguishing actor-related effects such as exfiltration of data or encryption of data, and defense-related effects, e.g.,~a preventive shutdown or whether passwords had to be reset.
We find that 5\% of all German universities have been impacted by a cyberattack between February 2022 and February 2023.

In addition, we provide an empirical analysis towards understanding the ecosystem of threat actors that specifically target German universities and academic institutions. 
Based on an analysis of a phishing activity cluster, we conclude with high confidence that some threat actors particularly focus on German academic institutions.
Over the course of more than three years, out of the 93~targeted universities, 58~(62\%) are German, representing a significant fraction with approximately 14\% of the total number of universities (423) in Germany.

\smallskip\noindent
To summarize, the main contributions of our work are:
\begin{enumerate}
\item We provide a timeline of cyberattacks against German universities, as confirmed by the targeted institutions or media reports, between February 2022 and February 2023, along with categorized impact.

\item We document a cluster of phishing attacks targeting universities, reconstructed through infrastructure and phishing email analysis, and trace these activities back to 2020. Moreover, we present a detailed description of the attackers’ \ac{TTP}, demonstrating that they exploit the infrastructures of multiple universities to disseminate malicious emails.

\item We propose countermeasures to prevent attacks, mitigate their impact, and implement proactive strategies that improve incident response capabilities.
\end{enumerate}

%% file: related-work.tex
\section{Related Work}
\label{sec:relwork}

Cyberattacks increasingly threaten higher education institutions worldwide, targeting them for financial gain or espionage.
While such attacks involve diverse techniques and objectives, phishing remains one of the most common initial access vectors used by threat actors to compromise university systems.
Prior work in this domain spans technical detection of phishing, analyses of user susceptibility, demographic profiling of potential targets, and sector-specific investigations into cyber threats facing academic institutions.
In the following, we review research and industry reports relevant to understanding phishing and cyberattacks in the higher education sector, with a particular focus on the context of our study.

\subsection{Phishing and Cyberattacks Against Higher Education Institutions}

There is growing evidence that higher education institutions are increasingly targeted by cyberattacks, including phishing campaigns.
Several industry reports document this trend.
Mandiant describes phishing campaigns specifically targeting at least 15 universities in the United States, underscoring academia’s attractiveness as a target due to the value of research data and intellectual property~\cite{MandiantHigherEd}.
Malwarebytes documents an 84\% increase in ransomware incidents targeting educational institutions over a six-month period in 2023, primarily driven by groups like Vice Society~\cite{MalwarebytesRansomware}.
Similarly, Asimily provides brief overviews of significant cyberattacks on \ac{U.S.} universities, highlighting both the frequency and impact of such incidents~\cite{AsimilyCyberattacks}.
Government reports support these observations.
According to the UK Cyber Security Breaches Survey 2025, 97\% of further and higher education institutions experienced phishing attacks, a significantly higher rate than businesses overall (85\%)~\cite{GovUKCyber}.
Moreover, higher education institutions were substantially more likely than businesses to report suffering negative outcomes from breaches.
Historic examples of targeted campaigns include the Silent Librarian attacks, during which Iranian actors systematically targeted universities to steal intellectual property and academic research data, as reported by the \ac{U.S.} Department of Justice~\cite {USDoJSilentLibrarian}.

In academic literature, a few studies have specifically investigated phishing in higher education.
The Alpha Phi-shing Fraternity study presents the results of a phishing simulation conducted at a higher education institution along with demographic data~\cite{AlphaPhiShing}.
A recent case study explores how university staff respond to phishing incidents, providing insights into incident response processes within educational institutions~\cite{PhishingIncidentResponseHE}.
Additionally, Masaryk University reports statistics on cyberattacks against its own institution, including over 121,000 attempted attacks in 2020, and discusses reasons why universities are valuable targets, though some conclusions remain anecdotal~\cite{MasarykCyberattacks}.

\subsection{Phishing Detection and Technical Approaches}

There are various detection and analysis techniques for addressing phishing threats.
For example, van der Heijden et al.\cite{cognitive_triaging} propose a machine learning-based approach to predict the likelihood of success for phishing emails, focusing on enhancing quantitative measurements of cognitive vulnerability triggers.
In another line of research, Ho et al.\cite{lateral_phishing} employ deep learning methods in which neural networks are trained on screenshots of phishing websites, leveraging visual features to detect phishing attempts.
Additionally, Ho et al.~\cite{credential_spearphishing} analyze user behavior and apply anomaly detection techniques to identify phishing threats.

\subsection{Human Factors and Target Demographics in Phishing Attacks}

Numerous studies have examined the human factors contributing to phishing susceptibility and sought to identify which user groups are most at risk of such attacks.
Research has explored psychological aspects influencing individuals’ ability to detect phishing, including threat perception and coping appraisals~\cite{ThreatCopingAppraisal}.
Surveys and experiments have assessed how different user demographics respond to phishing attempts, with several works focusing specifically on students in higher education contexts.
For example, Frauenstein reports on how 136 first-year students in South Africa interacted with phishing messages~\cite{StudentsResponses}.
Similarly, studies have found that international and first-year students are more likely to fall victim to phishing than domestic or senior students~\cite{PhishingRisksStudentCommunity}.
Other works have assessed phishing awareness and susceptibility among students in Nigeria~\cite{NigerianUniversityPhishing}, business students in the United States~\cite{UndergraduatesAtRisk}, and general university populations~\cite{AnalysisStudentVulnerabilities}.

Beyond higher education, broader efforts have analyzed large-scale telemetry data to empirically identify user segments at greater risk of email-based phishing and malware attacks~\cite{WhoIsTargeted}.
Complementing studies on user susceptibility, PhishInPatterns investigates how modern phishing websites employ sophisticated \ac{UX} and \ac{UI} design patterns to elicit user interactions and convey legitimacy, offering insights into user-facing aspects of phishing tactics at scale~\cite{PhishInPatterns}.

Despite important insights into user behavior and target profiling, many existing studies remain limited to small-scale surveys or narrow simulations and do not address the sector-wide threat landscape faced by higher education institutions.

\subsection{Summary and Research Gap}

While there exists substantial work on phishing detection and on user susceptibility, comparatively little research focuses explicitly on phishing and broader cyber threats against higher education institutions, particularly in the German context.
Much of the existing literature and reporting centers on universities in the United States or the United Kingdom, leaving a gap in understanding how such threats manifest in Germany.
Our work seeks to address this gap by analyzing phishing and cyberattack activities targeting universities in Germany, thereby contributing region-specific insights to the broader discourse on cybersecurity in higher education.

%% file: cctgu.tex
\section{Cyberattacks targeting universities in German-speaking regions}
\label{sec:attacks}
To contextualize this research, it is essential to provide an overview of confirmed cyberattacks that have targeted German-speaking academic institutions within a specified time frame. 
For data collection, we conduct a structured surface web research using predefined queries and publicly accessible sources such as news articles, press releases, and institutional statements. 
All identified incidents are cross-checked by at least two experienced cybersecurity researchers. 
Finally, we categorize the documented impact of each incident in six predefined categories to enable systematic comparison.

\subsection{Methodology}
Using German keywords in the context of cyberattacks against academic institutions, we employed web searches as a means to collect public reporting, either by the targeted organizations or via media reporting.
First, we used keyword searches using Google web search to find potential attacks, e.g., using the following phrases "cyberangriff hochschule", "cyberangriff universität", and "cyberattacke hochschule". 
After finding a potential victim, we used the victim's organization name and additional keywords, e.g., "phishing", "ransomware", or "ausfall", to collect more information on the incident, particularly to collect supplementary evidence on the incident, prioritizing institutional statements covering the incident where possible.
Finally, two experienced security researchers reviewed the collected information to assess the impact of the incident as described in more detail below.
We publish the resulting dataset, consisting of information shown in \Cref{tab:attacks} along with sources for each incident, referred to as \texttt{CCTGU23}\footnote{\url{https://osf.io/g57rj?view_only=481b15d000d54319aff0339de276234a}}.

\paragraph{Temporal scope}
We restrict our research to incidents that occurred between February 2022 and February 2023, both months included.
We focus on this time frame to ensure that information on occurred or suspected incidents is available, including those that were disclosed only months or even years after the actual attack. 
By selecting a period sufficiently in the past, we aim to capture a more complete and accurate picture of the threat landscape, allowing time not only for delayed disclosures but also for initial reports to be corroborated or falsified in cases where early information was inaccurate or based on rumor.

\paragraph{Regional filtering}
In this section, we restrict our research to universities and universities of applied sciences in the German-speaking regions of Germany, Austria, and Switzerland.
Although our study concentrates on Germany, our keyword-based methodology in the German language also yielded results pertaining to other German-speaking countries, specifically Austria and Switzerland.
We opted to incorporate these findings to present a more comprehensive and contextualized overview of the threat landscape.
On the other hand, we exclude incidents that may have happened in the considered time frame, but outside German-speaking areas, such as a ransomware attack targeting Université de Neuchâtel in February 2022~\cite{Neuchatel}.

\paragraph{Impact assessment}
For each incident, we categorize the impact using six aspects.
We consider two categories that are associated with actor behavior, i.e.,~exfiltration of data and encryption of data, as well as four incident response categories that describe defense behavior; in particular, preventive shutdown, uplink disconnect, password reset, and rebuild of systems.
We distinguish a preventive shutdown, which reflects turning off services completely so that they are not even reachable from within the organization, and an uplink disconnect, which is used to isolate an environment from the Internet, to prevent the actor from further command and control.
We rely on the information published by the affected institution or media reports, and mark a category even if only parts of the organization were affected.
For each incident, two experienced security researchers independently reviewed the available information and assigned an impact assessment. 
Discrepancies were discussed and resolved to ensure consistency. 
While financial data could enhance the assessment of incident impact, it is not included in this study because the affected universities did not disclose such information.
Similarly, we rarely found information on remediation procedures and their effectiveness, leading us to omit these aspects from our analysis.

\input{cctgu-table}

\subsection{Findings}
The resulting timeline highlights reported incidents and their impact. 
Note that the information presented here is based on publicly reported incidents, and additional, undisclosed attacks may have occurred during this period.
Despite scrutiny, it cannot be ruled out that a cyberattack matching our criteria may not have been captured in our dataset. 
Thus, we consider our dataset as a lower estimate.

For each institution in \Cref{tab:attacks}, the column \emph{Published} denotes when the attack was first published, either by the targeted organization or via media reporting.
If available, the column \emph{Attack time} contains the (approximate) point in time when the attack took place.
We do not distinguish between initial infection or any downstream phase in the cyber kill chain because most sources did not provide such information.
The column \emph{Ransom note?} denotes whether a ransom note was received.
The labels \emph{yes} and \emph{no} refer to cases in which the affected institutions confirmed whether a ransom note was received.
We consider cases as \emph{unknown} if the affected institution did not provide information.
We consider cases as \emph{unlikely} if no indication of a ransomware attack is given and the affected institution did not provide information.
We consider cases as \emph{likely} if the affected institution did not provide information and a ransomware actor claimed responsibility for the attack, e.g.,~via a dedicated leak site.
The six rightmost columns reflect impact categories as described above.

The resulting timeline is ordered chronologically and covers 25 incidents over the course of one year between February 2022 and February 2023.
The number of 25 confirmed incidents may appear low, especially considering that the Federal State Government of \ac{NRW}~\cite{NRWAnfrage} alone mentioned 77 affected institutions.
This difference in numbers is likely because the Federal State Government counted institutions and covers a five-year time span, while we only consider one year and focus on incidents instead of institutions. 
Furthermore, some attacks may not have been publicly described (yet included in the count by the Federal State Government) and thus are missing from our dataset.
In addition, we rule out attack attempts which have not shown any indication of attack impact, e.g., influx of incoming phishing mails, or lack any form of persistence, e.g., \ac{DDoS} attacks.

According to the Federal Statistical Office of Germany~\cite{GermanUniversities}, in 2022, there were 423 academic institutions in Germany, 20 of which (approx. 5\%) are included in \Cref{tab:attacks}.
Having five percent of all German universities affected within one year highlights that cyberattacks clearly impact the educational sector in Germany, especially considering that this only covers publicly documented incidents.
Most institutions (at least 24 out of~25) have suffered from an outage caused by the attack, i.e.,~preventive shutdown or uplink disconnect.
There is only one case where no impact on the availability was reported.

Given the prevalence of ransomware attacks, we tried to assess whether these incidents follow a ransomware scheme that would likely reflect a financial motivation of the actor. 
The structure and level of detail in the public statements or press releases vary considerably.
In only two cases, namely Universität Duisburg-Essen and \ac{HAW} Hamburg, the institutions publicly confirmed that a ransom note was received.
In both cases, an actor referred to as Vice Society claimed responsibility for the attacks via their dedicated leak site.
Similarly, Medizinische Universität Innsbruck is also mentioned on the leak site, indicating a ransom scheme, although not publicly confirmed by the affected institution.
In short, although only three incidents publicly indicate a ransomware scenario, by far most institutions experienced a serious impact.
Two further cases (\ac{FH} Westküste and \ac{HTW} Berlin) reported encryption of data but did not make a statement on whether a ransom note was received.
In addition, seven of the 25~victims confirmed that a ransom note has \emph{not} been received, and 13 cases did not provide information on a ransomware nexus.
This may indicate that the attack was not ransomware-driven, or the attack may have been mitigated before data was encrypted.
Alternatively, the actor may have considered the target as invaluable for financial reward.
To summarize, five of the 25~cases indicate a ransomware nexus.
However, while other motivations such as espionage or sabotage could be possible, we did not find any corroborating indication.
For the majority of attacks, the actor's intent cannot be inferred from the publicly available information.

In summary, while some universities publicly acknowledged having been targeted in a cyber attack, hardly any institution documents technical or financial impact, remediation procedures and their efficacy or even preventive measures in detail.
A notable exception is the remediation of the cyber attack on FH Münster in June 2022, which has been thoroughly documented in a talk by Ostkamp and Schinzel~\cite{schinzel2022youtube}.
We revisit lessons learned and aggregate recommendations in the conclusion of this work.

%% file: cctgu-table.tex
\begin{table*}[!tb]
    \centering
    \begin{footnotesize}
    \begin{tabular}{b{1.0cm}m{1.0cm}p{3.5cm}b{1.0cm}b{0.3cm}b{0.3cm}b{0.3cm}b{0.3cm}b{0.3cm}b{0.3cm}}
    \toprule
    \textbf{Attack \newline time} & \textbf{Published} & \textbf{Target} & \textbf{Ransom note?} & \rotatebox{90}{\textbf{exfiltration of data}} & \rotatebox{90}{\textbf{encryption of systems}} & \rotatebox{90}{\textbf{preventive shutdown}} & \rotatebox{90}{\textbf{uplink disconnect}} & \rotatebox{90}{\textbf{password reset}} & \rotatebox{90}{\textbf{rebuild of systems}}
    \tabularnewline \midrule[1pt]
    02/23 & 02/28/23 & TU Ilmenau & unlikely & & & \checkmark & \checkmark & & 
    \tabularnewline \midrule
    02/02/23 & 02/02/23 & Universität Zürich & no & & & & \checkmark & & \checkmark
    \tabularnewline \midrule
    02/03/23 & 02/05/23 & Universität Graz & no & & & \checkmark & & & 
    \tabularnewline \midrule
    ~ & 02/02/23 & Europäische Fachhochschule Rhein/Erft & no & & & \checkmark & & & 
    \tabularnewline \midrule
    ~ & 02/01/23 & Hochschule Ruhr-West & unlikely & & & \checkmark & & \checkmark & 
    \tabularnewline \midrule
    ~ & 02/01/23 & Hochschule Harz & unlikely & & & \checkmark & & &  \tabularnewline \midrule
    01/14/23 & 01/16/23 & Universität Innsbruck & no & & & & & \checkmark & 
    \tabularnewline \midrule
    01/16/23 & 01/20/23 & TU Bergakademie Freiberg & no & & & & \checkmark & & 
    \tabularnewline \midrule
    11/22 & 01/02/23 & FH Westküste & unknown & \checkmark & \checkmark & \checkmark & & \checkmark & 
    \tabularnewline \hline
    12/27/22 & 12/29/22 & Hochschule für Angewandte \newline Wissenschaften Hamburg & yes & \checkmark & \checkmark & \checkmark & \checkmark & \checkmark & \checkmark
    \tabularnewline \midrule
    12/23/22 & 12/30/22 & Westsächsische Hochschule \newline Zwickau & unknown & & & \checkmark & \checkmark & & \checkmark 
    \tabularnewline \midrule
    11/27/22 & 11/28/22 & Universität Duisburg-Essen & yes & \checkmark & \checkmark & \checkmark & \checkmark & \checkmark & \checkmark
    \tabularnewline \midrule
    11/12/22 & 11/12/22 & Technische Hochschule Ulm & unlikely & \checkmark & & & \checkmark & & 
    \tabularnewline \midrule
    ~ & 10/31/22 & Hochschule Heilbronn & unknown & & & \checkmark & \checkmark & \checkmark & \checkmark
    \tabularnewline \midrule
    10/19/22 & 10/25/22 & Hochschule für Technik, \newline Wirtschaft  und Kultur Leipzig & unlikely & & & & \checkmark & \checkmark & \checkmark
    \tabularnewline \midrule
    10/19/22 & 10/20/22 & Hochschule Ansbach & unlikely & \checkmark & & \checkmark & & \checkmark & 
    \tabularnewline \midrule
    ~ & 07/23/22 & Bergische Universität \newline Wuppertal & unlikely & & & \checkmark & & \checkmark & \checkmark 
    \tabularnewline \midrule
    06/18/22 & 06/21/22 & Medizinische Universität \newline \mbox{Innsbruck} & likely 
    & \checkmark & \checkmark & \checkmark & & \checkmark & \checkmark 
    \tabularnewline \midrule
    06/20/22 & 06/20/22 & FH Münster & no & \checkmark & & & \checkmark & \checkmark & 
    \tabularnewline \midrule
    06/06/22 & 06/08/22 & Pädagogische Hochschule \newline Freiburg & unknown & & & \checkmark & & & \checkmark
    \tabularnewline \midrule
    04/28/22 & 04/29/22 & Hochschule für Technik \newline und Wirtschaft Berlin & unknown & \checkmark & \checkmark & \checkmark & & & 
    \tabularnewline \midrule
    04/19/22 & 04/26/22 & Universitätsbibliothek Leipzig & unlikely & \checkmark & & \checkmark & & & 
    \tabularnewline \midrule
    03/27/22 & 04/05/22 & Paris Lodron Universität \newline Salzburg & no & & & \checkmark & & \checkmark & 
    \tabularnewline \midrule
    03/26/22 & 03/28/22 & TH Aschaffenburg & unknown & & & \checkmark & \checkmark & \checkmark & \checkmark 
    \tabularnewline \midrule
    02/24/22 & 02/25/22 & Hochschule Anhalt & unlikely & & & \checkmark & \checkmark & &
    \tabularnewline
    \bottomrule
    \end{tabular}
    \caption{Dataset \texttt{CCTGU23}: Confirmed cyberattacks targeting German-speaking universities between Feb.~2022 and Feb.~2023. This dataset is available as \ac{CSV} with sources.}
    \label{tab:attacks}
    \end{footnotesize}
\end{table*}

%% file: phishing-analysis.tex
\section{Tracking a cluster of phishing activity over three years}
\label{sec:phishing-analysis}
Although cyberattacks can exploit multiple infection vectors, this section concentrates specifically on credential phishing.
Compromised credentials obtained through phishing attacks facilitate a range of subsequent attack techniques, such as spear-phishing campaigns, account takeovers, and lateral phishing~\cite{lateral_phishing}, in which attackers leverage already compromised accounts within an organization to phish for additional internal credentials, thereby enabling adversaries to gain or expand initial access to targeted organizations.
According to IBM X-Force, phishing was employed to breach victim environments in approximately 25\% of incident response engagements in 2024~\cite{IBM2025}.
Similarly, Verizon’s 2025 Data Breach Investigations Report indicates that phishing accounted for initial access in 16\% of all breaches, excluding those caused by errors or misuse~\cite{Verizon2025}.
Both reports further highlight that credential abuse constitutes a substantial proportion of initial access incidents (30\% according to IBM and 22\% according to Verizon).
Notably, such credentials are often harvested through prior phishing campaigns, underscoring the critical need to understand and mitigate phishing threats.
As part of our analysis of phishing attacks targeting German academic institutions, we delve into a notable case study that provides valuable insights into the nature of these threats. 
In this case study, we examine a persistent phishing cluster for which attacks have been tracked over a span of three years, from 2020 to 2023. 
Triggered by incidents affecting our own institution, we conducted a technical analysis of the phishing emails and the associated infrastructure hosting the phishing forms.
The initial analysis indicated that the phishing emails appear to have been tailored to the target institution, and the hosting infrastructure allowed for pivoting to assess victimology. 
The decision to focus on this specific cluster is thus driven by its longevity and the wealth of victimology data available for analysis.
By tracking the activities of this specific cluster over an extended period, we aim to analyze the \acp{TTP} employed by the threat actor behind the cluster over time to uncover any persistent patterns in their modus operandi.
At the time of writing, the most recent activity of this cluster was observed in April 2026\footnote{April 16th 2026, targeting Hochschule Bochum~\url{https://urlscan.io/result/019d9612-6fc5-771a-8f74-312902fab08a}}.

\subsection{The helpdesk-themed phishing campaign}
\label{sub:helpdesk}
Over the course of a five-month period between September 2022 and February 2023, we tracked a phishing campaign targeting our university by collecting phishing emails addressed to staff and students in a series of four waves. 
Based on manual human inspection, we consider the activity to be part of an ongoing campaign based on the following observations:

\begin{enumerate}[label=(\alph*)]
    \item visually and technically similar credential phishing form,
    \item phishing site \ac{URL} pattern: using target-dependent subdirectories, often support- or helpdesk-themed, and
    \item a similar way to send the phishing mail: using credentials for one university to send to another, involving a third university's \ac{VPN} infrastructure.
\end{enumerate}

\begin{figure*}[tb]
 \centering
 \includegraphics[width=.8\textwidth]{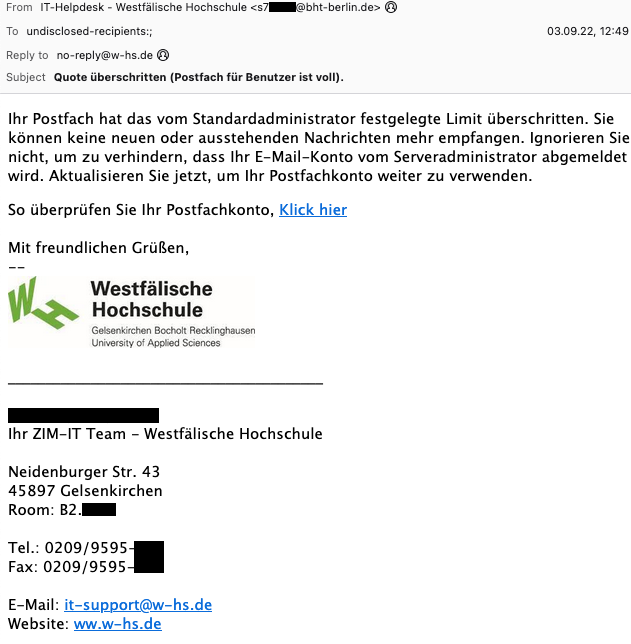}
 \caption{Screenshot of a phishing email as rendered in the Thunderbird mail user agent. The message was sent from a compromised account of the Berliner Hochschule für Technik. Written in German, it contains multiple linguistic errors. Its layout imitates legitimate IT‑Helpdesk emails of the Westphalian University of Applied Sciences, and the body urges the recipient to click a link to open the mail settings, claiming that the inbox quota has been exceeded. The link leads the recipient to a fraudulent webpage that harvests the victim’s university account credentials.}
 \label{fig:phishmail}
\end{figure*}

An example of a phishing mail is shown in \Cref{fig:phishmail}.
This email was received on 3 September 2022.
Following a typical phishing scheme, the email purports to notify about the mail quota being exceeded and prompts the user to visit the phishing site. 
The email exhibits the correct logo and signature text matching the target organization, i.e., they are tailored to the target organization.
The email text is in German and, based on incorrect grammar, most likely does not stem from a native German speaker, but rather appears to have been (machine-)translated into German.
All links in the email, even if the link text is meant to indicate otherwise, lead to the phishing site.

\begin{figure*}[tb]
 \centering
 \includegraphics[width=.8\textwidth]{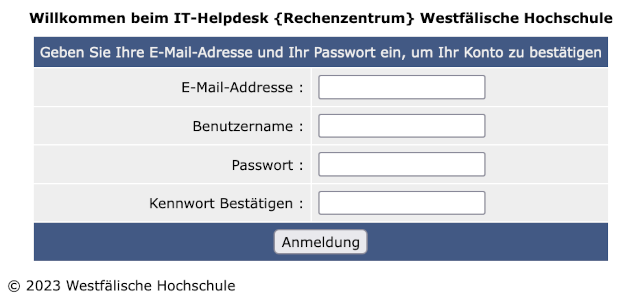}
 \caption{Screenshot of the phishing form hosted at {\footnotesize\texttt{hxxp[://]meethi[.]in/it-support[.]w-hs[.]de}}. The form requests the user's email address, username, password, and a repetition of the password. All text on the page is in German; the heading above the form reads “Welcome to the IT helpdesk of Westphalian University of Applied Sciences,” and the instruction within the form asks the user to enter their email address and password to verify the account. The submit button is labeled “sign‑in.” The page’s design is generic and does not resemble any official webpage of the Westphalian University of Applied Sciences.}
 \label{fig:phishsite}
\end{figure*}

In contrast to the tailored emails, the appearance of the credential phishing sites, as shown in \Cref{fig:phishsite}, is simple, and it rather looks like generic phishing without trying to mimic a specific, university-themed login interface.
Typically, the name of the targeted institution appears in the title of the phishing site, and occasionally in a copyright footer.

The \acp{URL} use a target-dependent path naming scheme.
For example, phishing sites targeting \texttt{w-hs.de} were hosted at the \acp{URL} shown in~\Cref{tab:urls}.

\begin{table*}[!tb]
    \centering
    \begin{tabular}{ll}
    \hline
        \textbf{Email received} & \textbf{Phishing URL} \\ \hline
        2022-09-03 & \texttt{immigrationlawyersinfo[.]com/\textcolor{red}{it-support.w-hs.de}} \\ 
        2022-10-15 & \texttt{patitasdelana[.]cl/\textcolor{red}{helpdesk.w-hs.de}} \\ 
        2022-12-20 & \texttt{meethi[.]in/\textcolor{red}{helpdesk.w-hs.de}} \\ 
        2023-02-19 & \texttt{meethi[.]in/\textcolor{red}{it-support.w-hs.de}} \\ \hline
    \end{tabular}
    \caption{\acp{URL} to the phishing sites indicate the targeted organization in the path}
    \label{tab:urls}
\end{table*}

We acknowledge that visual similarity alone may be considered insufficient to attribute phishing activity to a single threat actor, as templates may be shared, sold, or reused by multiple actors, e.g., in forums or via phishing kits. 
However, we provide additional supporting evidence that suggests a clustering of activity consistent with a single intrusion set, if not a singular actor.
First, we observe code artifact consistency.
Across the visually similar phishing pages targeting academic institutions, we identified a hidden form field named datacheck, filled with the value \texttt{memomomo@dos11.com}.
This field appears to serve as a rudimentary defense against credential poisoning or automated form submission, acting as a "flag" to emphasize some level of authenticity of data received from the phishing form.
Second, many of the observed phishing sites use a form submission action page ending in \texttt{HiTman2.php}.
Using web searches for both the specific datacheck values and the submission page name \texttt{HiTman2.php}, we found a limited set of sites that corroborate academic targeting, and we did not observe it in phishing pages outside the academic-themed cluster. 
Though the phishing kit appears partially generic, we found no publicly available kits or repositories (on GitHub or underground forums) that exhibited the exact code artifacts or visual appearance observed.
However, we cannot rule out that such a kit may be shared in private.

Conversely, since our analysis focuses on a single phishing page design, we cannot exclude the possibility that the threat actor may also employ alternative designs in related campaigns. 
To evaluate this likelihood, we performed spot checks by pivoting from the registered domains of several previously observed phishing sites, operating under the assumption that campaigns using different designs might share infrastructure. 
However, this investigation yielded no evidence of academic-themed phishing hosted on infrastructure overlapping with the domains under study.

\subsection{Clustering phishing attacks with a university nexus}
\label{sec:clustering}
\begin{figure*}[tb]
  \centering
  \includegraphics[width=\textwidth]{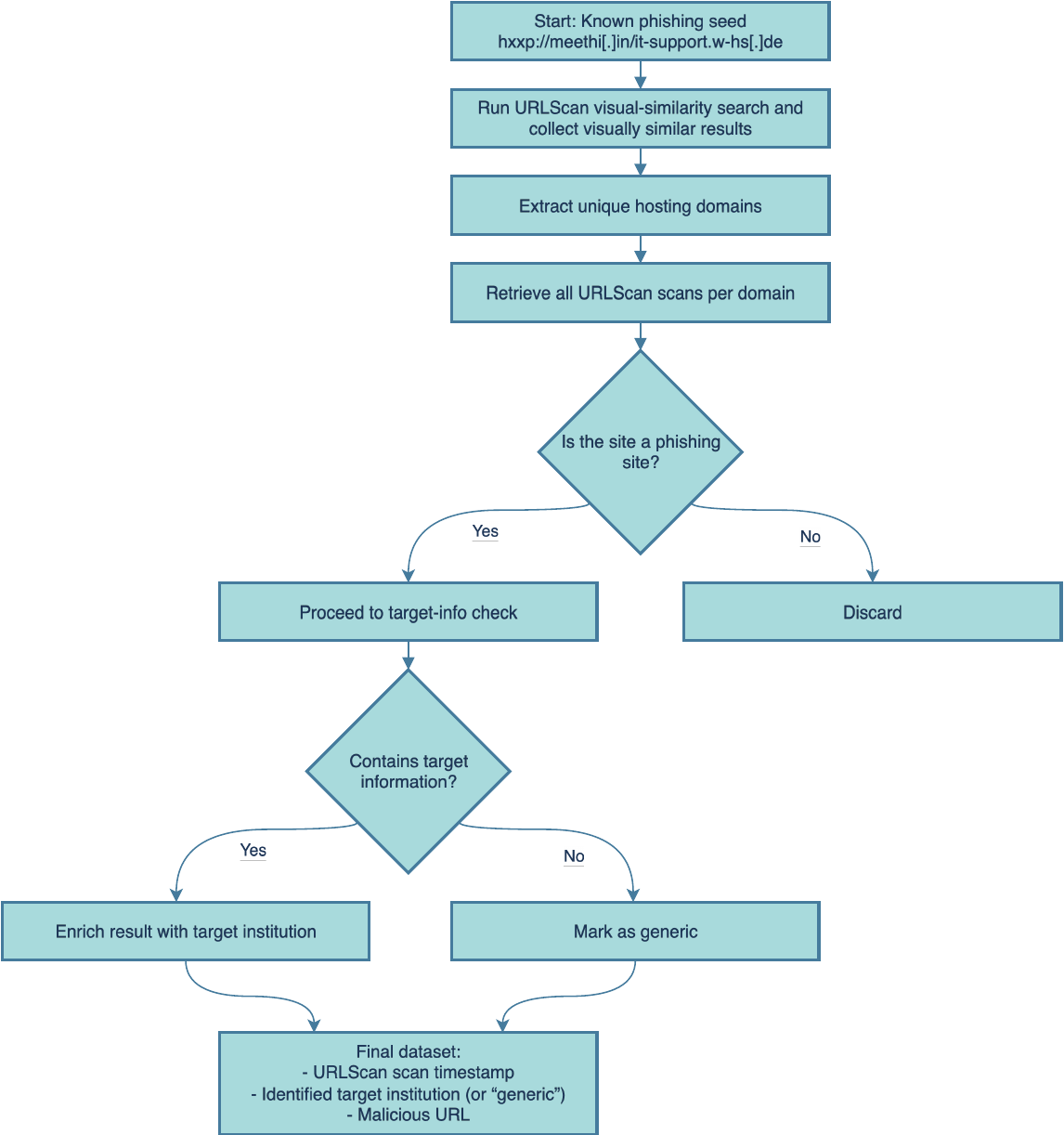}
  \caption{Procedure for pivoting from a reference phishing site to related sites using visual‑similarity and \ac{URL} analysis.}
  \label{fig:pivot}
\end{figure*}

To better understand the infrastructure and victimology of the phishing cluster, we used one of the known phishing sites as a pivot to identify additional sites belonging to the cluster.
Pivoting via the visual similarity of the rendered phishing sites and the observed hosting domains yields a set of phishing sites going back as far as 2020.
\Cref{fig:pivot} presents a schematic overview of our methodology for collecting additional sites belonging to the phishing cluster.
First, using the phishing site hosted at {\footnotesize \texttt{hxxp[://]meethi[.]in/it-support[.]w-hs[.]de}} as a reference, we employ URLScan's visual‑similarity search~\cite{UrlscanVisual} to identify scans that render pages with similar visual characteristics.
Secondly, we extract the domains from the URLScan results and retrieve all scans hosted on each of those domains.
From the resulting sites, we discard those that are not actual phishing pages because they lack a credential‑collecting form.
A defining characteristic of the phishing activity described in \cref{sub:helpdesk} is that the path component of the malicious \ac{URL} mimics the domain of the targeted institution. We exploit this property to filter the URLScan results, retaining only scans that disclose target information either in the URL path or within the phishing form.
Each site , in the resulting set of URLScan scans, is enriched with the likely targeted institution based on the \ac{URL} pattern, if present, or manually using the targeted institution's name from the website screenshot stored in URLScan.
In some cases, the phishing site does not indicate a particular target and seems generic, although it remains visually similar.
If the target cannot be inferred from the \ac{URL} or the screenshot, the scan is considered generic targeting.
The final dataset comprises, for each phishing site, the timestamp of the URLScan scan, the identified target institution, and the malicious \ac{URL}.
We publish the dataset spanning 442~scans as \texttt{PATU23}\footnote{Phishing attacks targeting universities (PATU23), password \texttt{infected}, \url{https://osf.io/7vbzs?view_only=481b15d000d54319aff0339de276234a}} and document our findings on the resulting cluster in this section.

\paragraph{Findings}
Until October 2023, using the pivoting described above, we identified a total of 270 phishing sites, of which 106 (39.3\%) do not exhibit a particular target and are considered as generic targeting.
The remaining 164 sites (60.7\%) exhibit 103 different targets, of which 93 (90\%) are universities.
Additionally, five institutions have been observed that are not universities but are associated with academia, namely two Fraunhofer institutions, Forschungszentrum Jülich, and two German university hospitals. 
Another five targets exhibit no academic association, precisely, a Mexican government institution, the Portuguese Guardia Nacional Republicana, an Italian political party, and two companies.
To summarize, 60\% of the sites reflect targeted activity, and 90\% of the targets are related to universities (95\% to academia in the broader sense).
This distribution highlights a particular focus of the educational sector by this actor.

The recovered data indicates that targeting of the educational sector dates back to March 2020.
Initially, between March and May 2020, Australian universities were in focus.
From May 2020 onward, the focus shifted to Europe, with German institutions accounting for the majority of targets.
Out of the 93 universities, 58~(62\%) are German, representing approximately 14\% of the total number of universities (423) in Germany, in 2023.
Australia (7 universities) and Italy~(6) follow, albeit with considerably lower frequencies.
In total, universities in 14 countries have been observed as targets of this cluster.
Given our focus on one cluster of phishing activity and a likely limited visibility due to the URLScan dataset, the fact that 14\% of all German universities were targeted with tailored phishing emails is significant.

\begin{figure*}[tb]
  \centering
  \includegraphics[width=.7\textwidth]{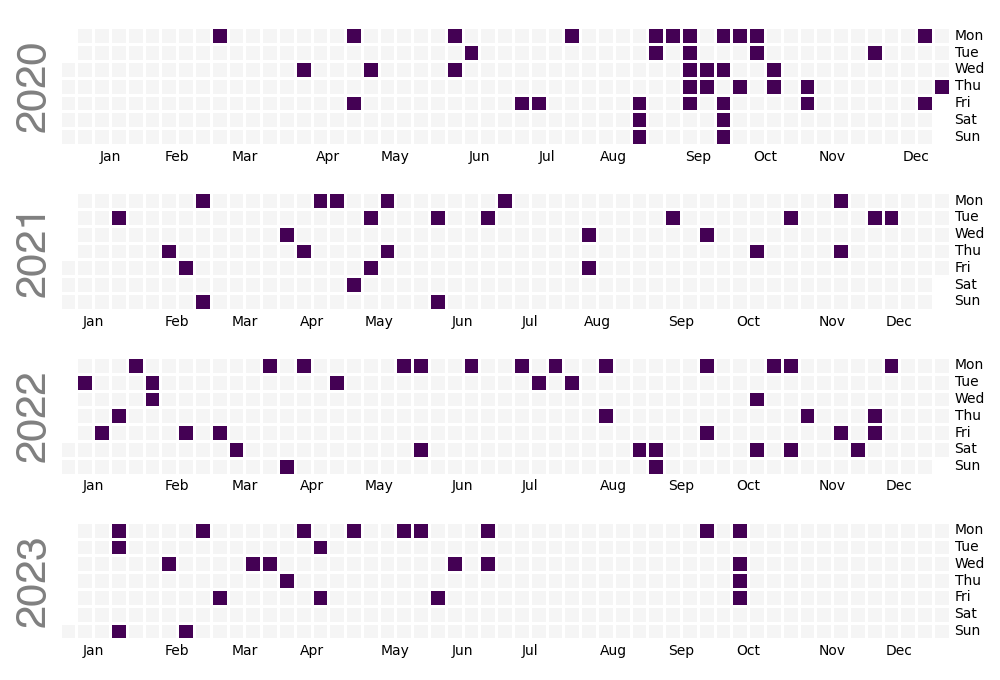}
  \caption{Calendar plot (February 2020 – October 2023) indicating the days on which a phishing site from the \texttt{PATU23} dataset was submitted to URLScan.}
  \label{fig:heatmap}
  %\vspace{-10pt}
\end{figure*}

Between March 2020 and June 2023, the activity can be observed on a regular basis with at least one new site per month.
\Cref{fig:heatmap} reflects points in time when a phishing site in the \texttt{PATU23} dataset has been submitted to URLScan. %, and yields a successful scan (i.e., the server returns HTTP status code 200).
Note that the timestamps do not directly reflect actor activity, but given that the phishing sites are often only active for a few days, we estimate that it corresponds roughly to the point in time when the institution was targeted by phishing.
In addition, while still significant, the resulting dataset likely only shows a subset of the actual phishing activity of this actor, given the limited visibility based on URLScan data.

\subsection{Infrastructure and modus operandi}
\label{sec:mo}

\begin{figure*}[tb]
  \centering
  \includegraphics[width=\textwidth]{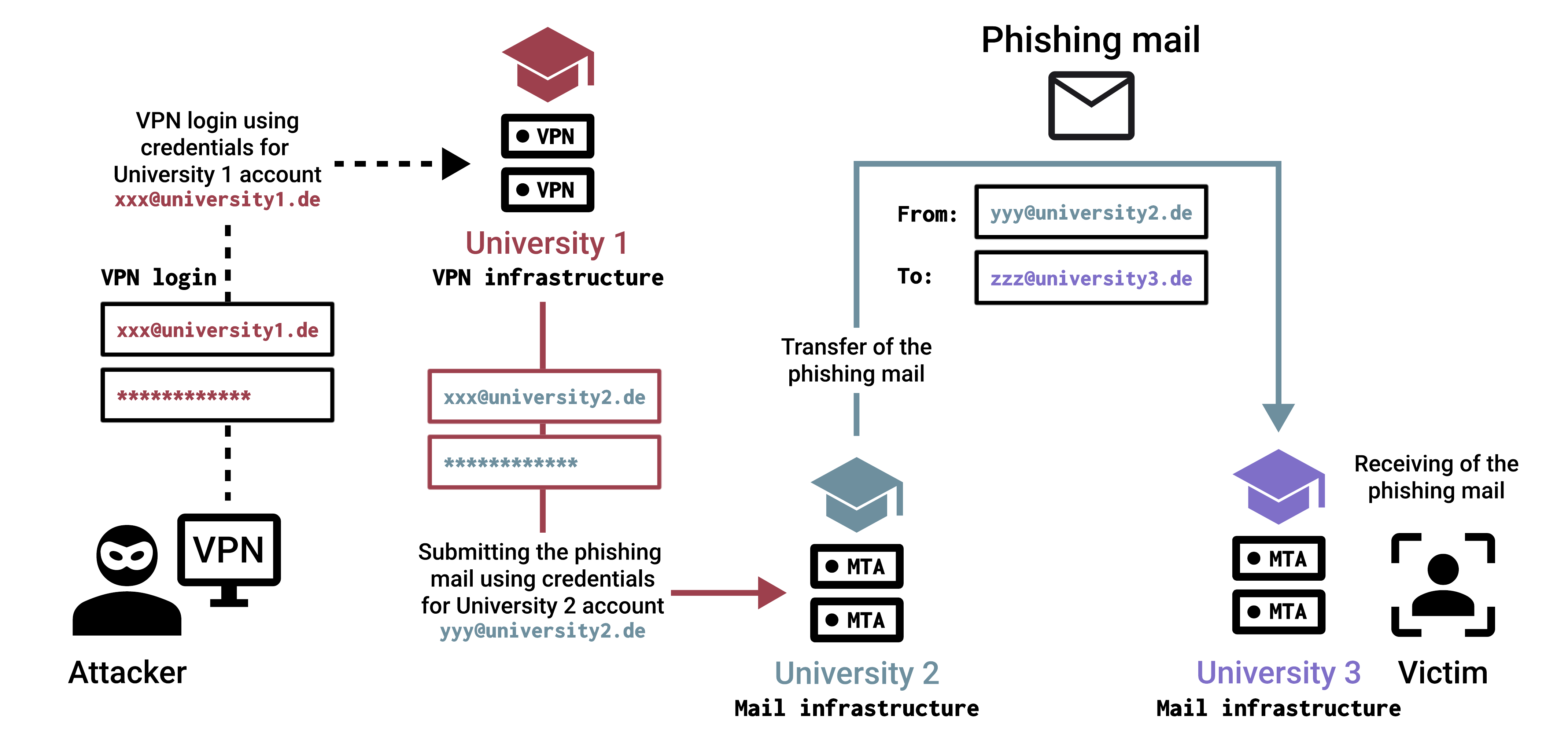}
  \caption{Procedure to send the phishing mail involving infrastructure from multiple academic institutions.}
  \label{fig:mo}
\end{figure*}

While the previous section documents the phishing emails and the phishing sites, this section looks at \emph{how} the emails are transmitted.
The actor behind this phishing activity exhibits a procedure, depicted in~\Cref{fig:mo}, where the emails are sent via university infrastructure, using valid credentials of university-associated personnel, likely compromised in previous phishing attacks.
First, the adversary uses valid \ac{VPN} credentials to log into the network of an academic institution, i.e., University~1.
Next, the adversary authenticates to the mail server of another academic institution, i.e., University~2, using valid credentials at University~2, to send the phishing emails to users at the target institution, i.e., University~3.
 
To confirm our findings about the mail path, we contacted a random subset of five other universities that appeared to have been targeted by the phishing activity based on the \texttt{PATU23} dataset.
We draw our conclusions from a total of 24~phishing emails involving 14~German academic institutions.
In all cases, the email's \textit{Received} header field, added by the \ac{MTA} of University~2, exhibits an academic \ac{VPN} \ac{IP} address used to submit the mail.

Clearly, the actor operates within the German-speaking educational sector, both on a \ac{VPN} level and with regard to the email sender credentials.
Possibly, this pattern is an attempt to mimic likely communication paths, intending to exploit inter-university trust.
In other words, it may be perceived as more likely for (German-speaking) academics to communicate amongst each other, even when \ac{VPN} access from a different institution may be involved.
As a result, the actor makes sure to only use credentials and infrastructure that are reliably associated with institutions in the educational sector.

\subsection{Actor motivation}
\label{sec:motivation}

Although we gained an understanding of the actor's modus operandi, the motivation is not clear. 
In this subsection, we attempt to shed light on \emph{why} the actor conducts the attacks.
By monitoring outgoing emails associated with accounts compromised by this actor, we observed two purposes:
Manual inspection of the subject header of each outgoing email allowed for either classifying it as a phishing email (indicated by a help desk or support topic) or as spam (indicated by donation and free money themes, as well as up to 1,000 recipients per email).
Assuming spam as the main monetization technique, the actor is likely financially motivated.
While we cannot verify that \emph{acquiring} the credentials and \emph{using} the credentials to send spam is conducted by the same actor, infrastructure overlap and timing support this hypothesis.

Bulk submission of the spam emails takes place during nighttime in the time zone associated with the compromised account, typically after 10~p.m. local time, possibly to increase chances of remaining undetected while tech staff is off duty.
In contrast, sending phishing emails typically occurs at daytime.
It is not limited to specific users but rather affects a significant fraction of staff and students. 
In one case, we observed 11\% of all staff and students as recipients of the phishing emails over the course of 14 days.
Some recipients have been targeted multiple times within the 14-day period.
While the phishing emails are sent exclusively via academic \acp{VPN} as described in \Cref{sec:mo}, the spam emails are submitted from both academic \ac{VPN} \ac{IP} addresses and \ac{IP} addresses not associated with academic institutions.

% Summarize findings of section 3

This section documents a prevalent phishing cluster targeting universities with a particular focus on German-speaking areas, affecting 14\% of all German universities.
The actor exhibits a procedure to involve university infrastructure when sending the phishing emails, most likely to cover their traces and reduce the chances of being detected.
We assess with moderate confidence that the actor is financially motivated based on the observed spam activities.

%% file: limitations.tex
\section{Limitations}
A limitation of the present study is its exclusive focus on German‑speaking universities, for which we have compiled a detailed catalogue of attacks (e.g., dates and impact categories).
Because no comparable dataset from non‑German institutions was assembled, we cannot assess whether the patterns observed are unique to the German higher‑education sector or reflect broader trends in academia.
Future work should therefore extend the analysis to universities in other countries, applying the same coding scheme and considering comparable time windows, to evaluate cross‑regional differences.

Our study does not provide complete coverage of the ecosystem of cyberattacks against academic institutions.
While we document one relevant phishing activity cluster targeting academic institutions in detail, there are further phishing activities, some of which use different \acp{TTP} and are likely from a different actor, although with similar targeting.
For instance, some universities have experienced phishing attacks using more targeted phishing forms, often mimicking the benign webmail interface of the targeted institution\footnote{Examples of tailored phishing forms include \ac{RWTH} Aachen \url{https://urlscan.io/result/c774ec45-dc65-416b-9cd6-7ececf09c78b} or Universität Hamburg \url{https://urlscan.io/result/35b61244-5da2-4baf-a8c9-bc76c8e98d33/}}.
However, given the difficulty in identifying and comparing activity clusters, compiling a dataset that reveals the targeting is non-trivial.
We focus on this specific cluster due to its longevity and the wealth of victimology data available for analysis.

Another limitation could be imposed by the nature of the URLScan dataset.
While scanning upwards of 100,000 URLs per 24 hours, we cannot rule out that the dataset may be skewed towards the security and infosec community, and possibly towards IT personnel in German-speaking academia.
This may have influenced the prevalence of German institutions in our dataset.
However, we are not aware of a service that better fits our requirements, especially providing visual similarity search over a large corpus of websites.

%% file: conclusion.tex
\section{Conclusion}
We showed that, with 25~confirmed cases over the course of 13~months, cyberattacks impact a significant fraction of universities in German-speaking areas, spanning at least 5\%~of all German universities.
In addition, we documented the victimology and \acp{TTP} of a phishing activity cluster spanning more than three years and showing a particular focus on universities in Germany. 
Out of the 93 targeted universities, 58 (62\%) are German, representing a significant fraction with approximately 14\% of the total number of universities (423) in Germany.
Thus, we conclude that German academic institutions have been targeted in particular.

Given that only four of the 25 confirmed cyberattacks exhibited a ransomware scenario, which in turn may indicate a financially motivated actor, it is difficult to infer the actor's motivation for the entirety of the attacks.
While the scale of the cyberattacks becomes measurable in our study, further research is needed to reliably determine the threat actor's motivation.
In particular, a financial motivation is questionable, given that universities in German-speaking areas are unlikely to pay, rendering a financial interest futile.
In contrast, in the cases we examined, we did not find any reliable indication of espionage or sabotage as motivation.

Going forward, it would be desirable to disclose more information on lessons learned during remediation and recovery from cyberattacks to assist future incident response, similar to the work of Ostkamp and Schinzel~\cite{schinzel2022youtube}.
Likewise, the disclosure of information on the financial impact of attacks and the suspected motivations of threat actors would be highly beneficial.
Finally, we aggregate recommendations on remediation and preventive measures in the context of cyberattacks targeting universities.

\paragraph{Remediation}

In the event of a successful cyberattack, universities should establish clear remediation procedures to minimize disruption and restore trust. 
A critical step is to enable rapid password resets for affected users. 
This requires not only technical readiness but also a reliable communication strategy to reach users through alternative channels, such as a pre-registered secondary email address or other verified backup communication means. 
Setting up a dedicated hotline can further facilitate user support during the crisis, ensuring timely guidance and response, as highlighted in the experiences reported by Ostkamp and Schinzel. 
In severe cases, a university-wide credential reset may be necessary to contain further compromise. 
To improve the speed and effectiveness of such measures, institutions should maintain up-to-date contact information for a crisis response team and ensure this team can be activated immediately.

\paragraph{Prevention}

To reduce the likelihood of future incidents, universities should adopt preventive measures that strengthen both technical defenses and organizational resilience.
Introducing \ac{2FA} for \ac{VPN} access significantly raises the barrier for attackers, as has been highlighted in prior incidents.
Applied to our observed phishing cluster in \cref{sec:mo}, \ac{2FA} would likely have rendered the attack strategy far more difficult for the attacker, if not entirely infeasible.
Similarly, the use of cryptographic email signatures has been discussed as a valuable lesson to reduce susceptibility to phishing.
The ransomware attack against FH Münster in June 2022 illustrates the importance of further safeguards: Ostkamp and Schinzel emphasized that maintaining offline backups is essential to ensure recovery in case critical systems are encrypted or destroyed. 
In the same context, they also underscored the importance of sustained investment in security awareness training for staff and students. 
Raising awareness about phishing tactics and basic cyber hygiene can substantially limit the effectiveness of initial compromise attempts and foster a more resilient academic environment.

\begin{acks}
This work received support from the German Federal Ministry of Research, Technology and Space (BMFTR) (grants 13FH101KB1 and 16KIS1746), and nicos AG.
The authors gratefully acknowledge the support of \url{urlscan.io}, which provided access to their platform, facilitating the comprehensive data acquisition required for this study.
\end{acks}